\documentclass[%
aps,twocolumn,PRE,reprint,superscriptaddress,
amsmath,amssymb,
]{revtex4-2}

\usepackage{hyperref}
\usepackage{graphicx}
\usepackage{dcolumn}
\usepackage{bm}

\usepackage{algorithm}
\usepackage{algpseudocode}

\begin{document}
\bibliographystyle{apsrev4-2}

\title{Social norms and cooperation in higher-order networks}

\author{Yin-Jie Ma}
\affiliation{School of Business, East China University of Science and Technology, Shanghai 200237, China}
\affiliation{Research Center for Econophysics, East China University of Science and Technology, Shanghai 200237, China}
\affiliation{CNR - Institute of Complex Systems, Via Madonna del Piano 10, I-50019 Sesto Fiorentino, Italy}

\author{Zhi-Qiang Jiang}
\email{zqjiang@ecust.edu.cn}
\affiliation{School of Business, East China University of Science and Technology, Shanghai 200237, China}
\affiliation{Research Center for Econophysics, East China University of Science and Technology, Shanghai 200237, China}

\author{Fan-Shu Fang}
\affiliation{CNR - Institute of Complex Systems, Via Madonna del Piano 10, I-50019 Sesto Fiorentino, Italy}
\affiliation{College of Economics and Management, Nanjing University of Aeronautics and Astronautics, Nanjing 211101, China}

\author{Matja{\v z} Perc}
\affiliation{Faculty of Natural Sciences and Mathematics, University of Maribor, Koro{\v s}ka cesta 160, 2000 Maribor, Slovenia}
\affiliation{Department of Medical Research, China Medical University Hospital, China Medical University, Taichung 404332, Taiwan}
\affiliation{Complexity Science Hub Vienna, Josefst{\"a}dterstra{\ss}e 39, 1080 Vienna, Austria}
\affiliation{Department of Physics, Kyung Hee University, 26 Kyungheedae-ro, Dongdaemun-gu, Seoul, Republic of Korea}

\author{Stefano Boccaletti}
\email{stefano.boccaletti@gmail.com}
\affiliation{CNR - Institute of Complex Systems, Via Madonna del Piano 10, I-50019 Sesto Fiorentino, Italy}
\affiliation{Sino-Europe Complexity Science Center, School of Mathematics, North University of China, Taiyuan 030051, China}
\affiliation{Research Institute of Interdisciplinary Intelligent Science, Ningbo University of Technology, Zhejiang, Ningbo 315104, China}

\begin{abstract}
Recent research has focused on understanding how cooperation is fostered through various mechanisms in cognitive settings, particularly through pairwise interactions. However, real-world interactions often extend beyond simple dyads, including multiple cliques with both pairwise and higher-order interactions. These complex interactions influence how individuals perceive and adapt their strategies based on social norms. We here introduce a model that explores the evolution of collective strategies and social norms within a heterogeneous environment, encompassing both dyadic and three-body interactions. We find that social norms play a crucial role in promoting cooperation in comparison to simply imitating the most successful neighbor. We also show that the rise of prosocial norms leads to increased cooperation across various social dilemmas, often resulting in shifts from defective to cooperative behavior. Additionally, we observe that a moderate level of information privacy helps sustaining prosocial norms and curtails antisocial tendencies, even in situations where mutual defection might seem advantageous. Our research thus offers insights into the evolution of cooperation through the lens of social norm diffusion in higher-order networks.
\end{abstract}

\date{\today}

\maketitle
\section{Introduction}\label{Sec:Introduction}
Cooperation, where individuals willingly bear costs to benefit others, is a widespread phenomenon across various biological and social spheres, as highlighted in numerous studies \cite{Pisauro-Fouragnan-Arabadzhiyska-Apps-Philiastides-2022-NatCommun, Norenzayan-Shariff-2008-Science, Yang-Zhang-Charness-Li-Lien-2018-ProcNatlAcadSciUSA, Ye-Zhao-Zhou-2022-ChaosSolitonsFractals, Cimpeanu-DiStefano-Perret-Han-2023-ChaosSolitonsFractals, Dhakal-Chiong-Chica-Han-2022-RSocOpenSci,Shu-Fu-2023-ProcRSocA-MathPhysEngSci}. The subject is so significant that it has been listed as one of the top 25 scientific puzzles by the renowned journal \textit{Science} in 2005 \cite{Kennedy-Norman-2005-Science}. In-depth investigations into diverse ecological factors, such as memory \cite{Park-Nowak-Hilbe-2022-NatCommun, Hilbe-MartinezVaquero-Chatterjee-Nowak-2017-ProcNatlAcadSciUSA}, reputation \cite{Fu-Hauert-Nowak-Wang-2008-PhysRevE, Gross-DeDreu-2019-NatCommun}, repeated interaction \cite{Pacheco-Traulsen-Ohtsuki-Nowak-2008-JTheorBiol, VanVeelen-Garcia-Rand-Nowak-2012-ProcNatlAcadSciUSA}, and network structure \cite{Santos-Pacheco-2005-PhysRevLett,GomezGardenes-Campillo-Floria-Moreno-2007-PhysRevLett,Fu-Hauert-Nowak-Wang-2008-PhysRevE,Santos-Santos-Pacheco-2008-Nature,Floria-GraciaLazaro-GomezGardenes-Moreno-2009-PhysRevE,Fu-Wang-Nowak-Hauert-2009-PhysRevE,GomezGardenes-GraciaLazaro-Floria-Moreno-2012-PhysRevE,Allen-Lippner-Chen-Fotouhi-Momeni-Yau-Nowak-2017-Nature,Fotouhi-Momeni-Allen-Nowak-2019-JRSocInterface}, have shown that cooperation can be more beneficial than selfish behavior in certain situations. Building on the groundbreaking work of Nowak \cite{Nowak-2006-Science}, five fundamental mechanisms - kin selection, direct reciprocity, indirect reciprocity, group selection, and network reciprocity - have been established to systematically understand and explain the rise of cooperative behavior. Significant research efforts have been focused on refining these mechanisms, especially indirect reciprocity. This includes enhancing our understanding of cognitive processes \cite{Leimar-Hammerstein-2001-ProcRSocB-BiolSci, Schmid-Chatterjee-Hilbe-Nowak-2021-NatHumBehav} and modeling realistic reciprocal actions within various social norms \cite{Schmid-Ekbatani-Hilbe-Chatterjee-2023-NatCommun}.

Social norms play a crucial role in shaping how individuals assess others' reputations and adjust their strategies accordingly \cite{Chen-Wang-Fu-2022-NewJPhys, Han-2022-JRSocInterface, Xia-Wang-Perc-Wang-2023-PhysLifeRev, Chen-Fu-2023-PNASNexus}. The process of evaluation can be categorized into four distinct levels based on the combination of the donor's and recipient's strategies and reputations. The simplest, first-order evaluation, focuses solely on the donor's strategy. In this scenario, image scoring has proven effective for identifying cooperators and penalizing defectors \cite{Nowak-Sigmund-1998-Nature}. The second-order evaluation expands this by considering both the donor's strategy and the recipient's reputation, with dominant strategies including stern judging and simple standing \cite{Santos-Santos-Pacheco-2018-Nature}. The third-order evaluation adds another layer by taking into account the donor's reputation and strategy, along with the recipient's reputation, but not their strategy \cite{Ohtsuki-Iwasa-2006-JTheorBiol}. The most complex, fourth-order evaluation, simultaneously looks at both the strategy and reputation of the donor and the recipient \cite{Brandt-Sigmund-2004-JTheorBiol}.

However, these rules typically assume that reputation information is widely known and trusted upon, which contrasts starkly with real-world situations characterized by noise, differing opinions, and imperfect information \cite{Hilbe-Schmid-Tkadlec-Chatterjee-Nowak-2018-ProcNatlAcadSciUSA, Han-Perret-Powers-2021-CognSystRes}. Even in cases of complete public information without distortion, third-party evaluations and interactions among neighbors vary depending on the individual's social network and the specific social norm they adhere to. This variability significantly affects reputation assessments through various mechanisms such as gossip \cite{Sommerfeld-Krambeck-Semmann-Milinski-2007-ProcNatlAcadSciUSA}, conformity \cite{Huang-Li-Jiang-2023-PhysRevE}, peer pressure \cite{Gao-Pan-He-2023-ChaosSolitonsFractals}, and especially through collective strategies or opinion aggregation in group decisions \cite{Civilini-Anbarci-Latora-2021-PhysRevLett}.


To better understand the mechanisms that encourage cooperation and to thoroughly explore the impact of social norms derived from collective strategies (specifically, how neighbors assess an individual) in various neighborhoods, two approaches have been instrumental: evolutionary game theory (EGT) and simplicial complexes. EGT, in one respect, is epitomized by a generalized two-strategy game involving two participants \cite{Perc-Jordan-Rand-Wang-Boccaletti-Szolnoki-2017-PhysRep}. In this game, each player independently decides to either cooperate (C) or defect (D), without prior knowledge of the other's choice. The outcomes of these strategies are represented in a payoff matrix with variables $R$ (reward), $S$ (sucker's payoff), $T$ (temptation to defect), and $P$ (punishment), each reflecting different combinations of the players' strategies \cite{Tanimoto-2007-PhysRevE}.

In contrast, simplicial complexes expand the scope of EGT by incorporating higher-order interactions \cite{AlvarezRodriguez-Battiston-deArruda-Moreno-Perc-Latora-2021-NatHumBehav,Guo-Jia-SendinaNadal-Zhang-Wang-Li-AlfaroBittner-Moreno-Boccaletti-2021-ChaosSolitonsFractals,Wan-Ichinose-Small-Sayama-Moreno-Cheng-2022-ChaosSolitonsFractals,Fan-Yin-Xia-Perc-2022-ProcRSocA-MathPhysEngSci}, enabling the analysis of collective strategies adopted by various groups. As a comprehensive model that includes both pairwise and higher-order interactions \cite{Battiston-Cencetti-Iacopini-Latora-Lucas-Patania-Young-Petri-2020-PhysRep,Battiston-Amico-Barrat-Bianconi-deArruda-Franceschiello-Iacopini-Kefi-Latora-Moreno-Murray-Peixoto-Vaccarino-Petri-2021-NatPhys,Majhi-Perc-Ghosh-2022-JRSocInterface}, simplicial complexes effectively integrate various types of interactions, particularly those within groups, into the structure of social networks. This approach is ideal for examining how social norms and collective strategies within diverse neighborhoods and interactions may facilitate cooperation across different dilemmas.

In our study, we consider four distinct social norms, which are: “kick Jimmy when he is down” (KJD, choosing to defect when the overall strategy in the neighborhood is defection), “send in the cavalry” (SIC, opting to cooperate when the neighborhood's collective strategy is defection), “add the icing on the cake” (AIC, cooperating when the collective strategy is cooperation), and “envy breeds hate” (EBH, defecting when the collective strategy is cooperation). These norms have significant psychological and societal implications. KJD, for instance, embodies the concept of imposing additional penalties on those inclined to defect or free-ride \cite{Rockenbach-Milinski-2006-Nature}. SIC is indicative of offering a second chance \cite{Shehada-Yeun-Zemerly-AlQutayri-AlHammadi-Hu-2018-JNetwComputAppl} or demonstrating unwavering loyalty \cite{Fu-Guo-Cheng-Huang-Chen-2019-PhysicaA} towards those with a tarnished reputation. AIC reflects a tendency to support close individuals with positive reputations \cite{Wang-Chen-Wu-2021-ApplMathComput}, while EBH captures the tendency to act against those held in high esteem due to envy \cite{Szolnoki-Xie-Ye-Perc-2013-PhysRevE}.

These social norms effectively encompass all individual strategies within a given collective neighborhood context. To enhance the understanding of this topic, we integrate these norms with 1-simplex (pairwise interactions via a link) and 2-simplex (three-body interactions via a triangle). This approach allows us to explore how collective strategies are influenced by a combination of pairwise and three-body interactions in diverse neighborhoods. Additionally, we examine the reasons these social norms either facilitate or impede the development of cooperation in various game scenarios.

The remainder of our paper is organized into the following sections. In Section \ref{Sec:Model}, we delve into the configurations of evolutionary games, providing clear definitions for both collective strategy and social norms within the context of simplicial complexes. This section covers a range of models: Model I, which uses a simple best imitation approach; Model II, which broadens the scope to include a variety of social norms; and Model III, an innovative model that introduces social norms within a noisy environment characterized by private information. Section \ref{Sec:Results} presents detailed phase diagrams for these models, exploring how they behave under different game scenarios and varying levels of noise and information privacy. This section also includes data on the prevalence of different social norms, shedding light on how cooperation emerges in these contexts. Finally, in Section \ref{Sec:Conclusions}, we provide a summary of our main findings and discuss potential directions for further research building on our work.

\section{Model formulation}\label{Sec:Model}

\subsection{Preliminaries of simplicial complexes and EGT}\label{Sec:Preliminaries}
In this section, we introduce preliminary settings to integrate different games with the evolution of cooperation on simplicial complexes. Without prejudice to generality, we assume that each player $i$ in population $N$ is initially assigned to a strategy set $S_i = \left\{s_{ij} | j \in N_i, s_{ij} \in \{0,1\} \right\}$ at round $t = 0$, where $N_i$ is the neighbourhood of player $i$ and $s_{ij} = 1$ ($s_{ij} = 0$) if $i$ cooperates (defects) with his neighbor $j$. $k_i$ is the degree of player $i$. In the very beginning of the evolutionary process $t = 0$, we randomly assign a social norm $\phi_i$ to each player $i$. $\Phi = \left\{\phi_{i} | i \in N, \phi_{i} \in \{1, 2, 3, 4\} \right\}$ denotes each player's social norm and controls how they evaluate their neighbors and accordingly changes his own $S_i$ in the next round $t + 1$. Each $s_{ij}$ and $\phi_i$ are initialized with a random value selected uniformly from their respective optional sets at $t = 0$.

\begin{figure}
\centering
\includegraphics[width=1\linewidth]{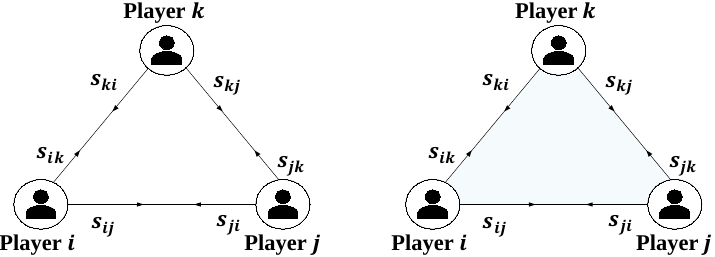}
\caption{\label{Fig:Triangles}Sketches of pairwise interactions in a clique (left panel) and three-body interactions in a 2-simplex (right panel). $1 - \rho$ fractions of triangles in the network will be formed in the type of left panel while $\rho$ fractions in the opposite.}
\end{figure}

In order to investigate the evolution of individual interactions under different social norms on simplicial complexes, we firstly generate a simplicial complex among $N = 300$ players. Following Ref.\cite{Kovalenko-SendinaNadal-Khalil-Dainiak-Musatov-Raigorodskii-AlfaroBittner-Barzel-Boccaletti-2021-CommunPhys}, a fully connected subgraph $G_0$ is generated with $N_0 = 5$. After that, $m = 1$ new player is added to $G_0$, who is linked to the two players of randomly-chosen $m$ edge that already exists in $G_0$ and thus formulates $m$ new triangles. With the repeating steps of adding $m$ edges to $G_0$, a network $G$ is formed until all the $N$ players are in $G$. To distinguish each triangle from a 2-simplex and a closure of 1-simplices, we utilize a parameter $\rho \in [0, 1]$ and regulate that $\rho$ fraction of triangles are considered as 2-simplices and $1 - \rho$ as the opposite \cite{Guo-Jia-SendinaNadal-Zhang-Wang-Li-AlfaroBittner-Moreno-Boccaletti-2021-ChaosSolitonsFractals}. Fig.~\ref{Fig:Triangles} illustrates the differences of these two structures. We denote $C^2_i \subseteq N_i$ as neighbors who are in at least one common 2-simplex with player $i$ in $N_i$ and $C^2_{ij} = C^2_i \cap C^2_j$ as player $i$ and player $j$'s common 2-simplex neighbors.

\begin{figure}
\centering
\includegraphics[width=0.6\linewidth]{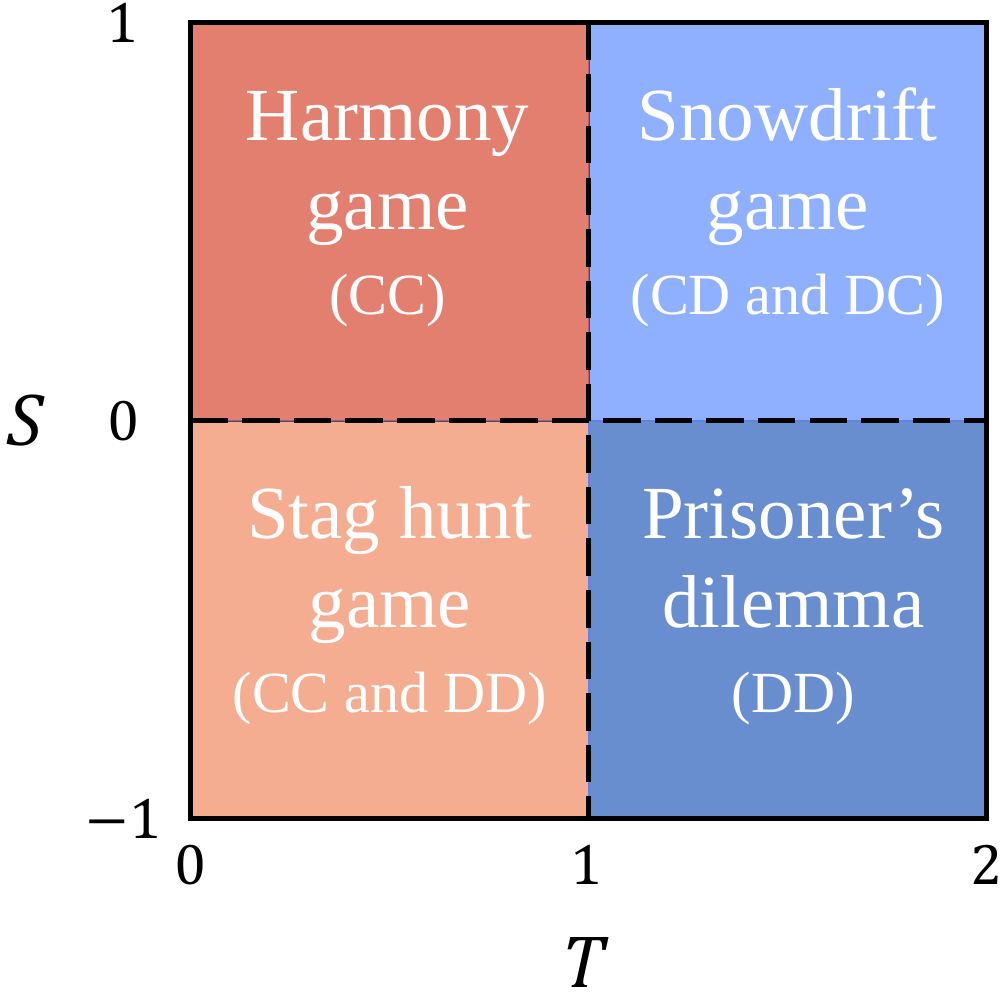}
\caption{\label{Fig:SocialDilemmas}Illustrations of typical pairwise games. When reward ($R$) and punishment ($P$) are fixed to be 1 and 0, respectively, $T-S$ space categorizes these games into four quarters: Harmony game, Stag hunt game, Snowdrift game, and Prisoner's dilemma. Notations in brackets represent the corresponding Nash equilibria in different games, where C is cooperation and D represents defection.}
\end{figure}

In terms of the payoffs during the evolutionary dynamics, we use a typical matrix to depict all the potential results when reward $R = 1$ and punishment $P = 0$:
\begin{equation}\label{Eq:Game:PayoffMatrix}
    \bordermatrix{%
      & C & D\cr
    C & 1 & S\cr
    D & T & 0
    }
\end{equation}
Notably, a pairwise game is depicted by only two parameters, $T$ and $S$ ($0 \leq T \leq 2, -1 \leq S \leq 1$). As is fully discussed in Ref.\cite{Guo-Jia-SendinaNadal-Zhang-Wang-Li-AlfaroBittner-Moreno-Boccaletti-2021-ChaosSolitonsFractals, Xia-Wang-Perc-Wang-2023-PhysLifeRev}, Matrix \ref{Eq:Game:PayoffMatrix} can be divided into four different categories with respect to the combinations of these two parameters: Harmony game (H), Stag hunt game (SH), Snowdrift game (SD), and Prisoner's dilemma (PD). According to the $T-S$ space of these dilemmas, Nash equilibria differ when $T$ and $S$ change (see Fig.~\ref{Fig:SocialDilemmas}).

\subsection{Model I: A benchmark model}\label{Sec:BenchmarkModel}

In each round of our benchmark model, every player $i$ experiences 2 steps: (1) calculate payoff $\Pi_i$ on simplicial complexes, and (2) change strategy $S_i$ towards neighbors. $\Pi_i$ is accumulated by each single payoff $\Pi_{i, (ij)}$ on the specific link $(i, j)$:
\begin{equation}\label{Eq:PayoffForPlayeri}
	\Pi_i = \frac{1}{k_i} \sum_{j \in N_i} \Pi_{i, (ij)},
\end{equation}
where $\Pi_{i, (ij)}$ is the aggregation which iterates all the elementary $\tau$ in the set of $\Delta$ that contains all the $k_{ij}$ triangles formed by the link $(i, j)$. $\Pi_{i, (ij), \tau}$ is the payoff of player $i$ with respect to link $(i, j)$ in the specific triangle $\tau$:
\begin{equation}\label{Eq:PayoffForLinkiij}
	\Pi_{i, (ij)} = \frac{1}{k_{ij}} \sum_{\tau \in \Delta} \Pi_{i, (ij), \tau}
\end{equation}
If the triangle $\tau$ is a closure of 1-simplices, link $(i, j)$ is nothing but a classic pairwise game (Game 1) and $\Pi_{i, (ij), \tau}$ can be obtained by the following payoff matrix:
\begin{equation}\label{Eq:Game1:PayoffMatrix}
    \bordermatrix{%
      & C & D\cr
    C & 1 & S_1\cr
    D & T_1 & 0
    }
\end{equation}
Otherwise, if $\tau$ is a 2-simplex where link $(i, j)$ and $\Pi_{i, (ij), \tau}$ should be entangled with the third parties, two circumstances are listed according to the potential combinations of $s_{ki}$ and $s_{kj}$:\\
1. If $s_{ki} = s_{kj}$, then, player $i$ and $j$ follow Game 2 and gain payoff from the matrix:
\begin{equation}\label{Eq:Game2:PayoffMatrix}
    \bordermatrix{%
      & C & D\cr
    C & 1 & S_2\cr
    D & T_2 & 0
    }
\end{equation}
2. If $s_{ki} \neq s_{kj}$, then, player $i$ and $j$ follow Game 3 and gain payoff from the matrix:
\begin{equation}\label{Eq:Game3:PayoffMatrix}
    \bordermatrix{%
      & C & D\cr
    C & 1 & S_3\cr
    D & T_3 & 0
    }
\end{equation}

An evolutionary step in round $t$ goes forward to simultaneously change strategy $S_i$ towards neighbors after every player $i$ calculates his own payoff. Before introducing the updating mechanism of $S_i$, we first denote $s'_{ij}$ as a collective strategy towards player $j$ from the view of player $i$, and $s^*_{ij}$ as a potential strategy that player $i$ will take to enhance his payoff:

\begin{equation}\label{Eq:CollectiveStrategy}
    s'_{ij} =
    \begin{cases}
    1, & \text{if } \frac{1}{k^C_{ij}} \sum_{k \in C^2_{ij}} s_{kj} \geq \theta \text{ and } C^2_{ij} \neq \emptyset \\
    0, & \text{if } \frac{1}{k^C_{ij}} \sum_{k \in C^2_{ij}} s_{kj} < \theta \text{ and } C^2_{ij} \neq \emptyset \\
    \emptyset, & \text{if } C^2_{ij} = \emptyset
    \end{cases}
\end{equation}

\begin{equation}\label{Eq:Potentialstrategy}
    s^*_{ij} =
    \begin{cases}
    s_{j^{*}i}, & \text{with probability } \frac{1} {{1 + \exp[(\Pi_i - \Pi_{j^{*}})/\kappa]}} \\
    s_{ij}, & \text{otherwise}
    \end{cases}
\end{equation}
where $k^C_{ij}$ is the number of common players in $C^2_{ij}$ and $\theta$ controls how collective strategies are integrated. $j^{*}$ is the player with the highest payoff in the neighbourhood $N_i$. Here we set $\theta = 0.5$ to apply the rules of majority while two-third majority and proportion majority can be tested as well \cite{Hastie-Kameda-2005-PsycholRev, Civilini-Anbarci-Latora-2021-PhysRevLett}.

Then each player $i$ simultaneously updates all of his strategies $s_{ij}$ towards neighbors. In our benchmark model, every player updates his strategy $s_{ij}$ with $s^*_{ij}$ no matter collective strategies $s'_{ij}$ is to cooperate with player $j$ ($s'_{ij} = 1$), defect him ($s'_{ij} = 0$) or there is no collective strategy for reference ($s'_{ij} = \emptyset$). In other words, social norm and collective strategy will not influence the updates of $s_{ij}$ in Model I and, therefore, every player imitates $s^*_{ij}$ according to Eq.~\ref{Eq:Potentialstrategy}.

An evolutionary round ends when every player calculates his payoff and updates strategy $s_{ij}$. After that, $t \leftarrow t + 1$ and players repeat the same procedures until fractions of different $s_{ij}$ stabilize, when an iteration of the evolutionary process comes to an end.

\subsection{Model II: Evolutionary games with different social norms}\label{Sec:SNModel}
\begin{figure*}
\centering
\includegraphics[width=0.95\linewidth]{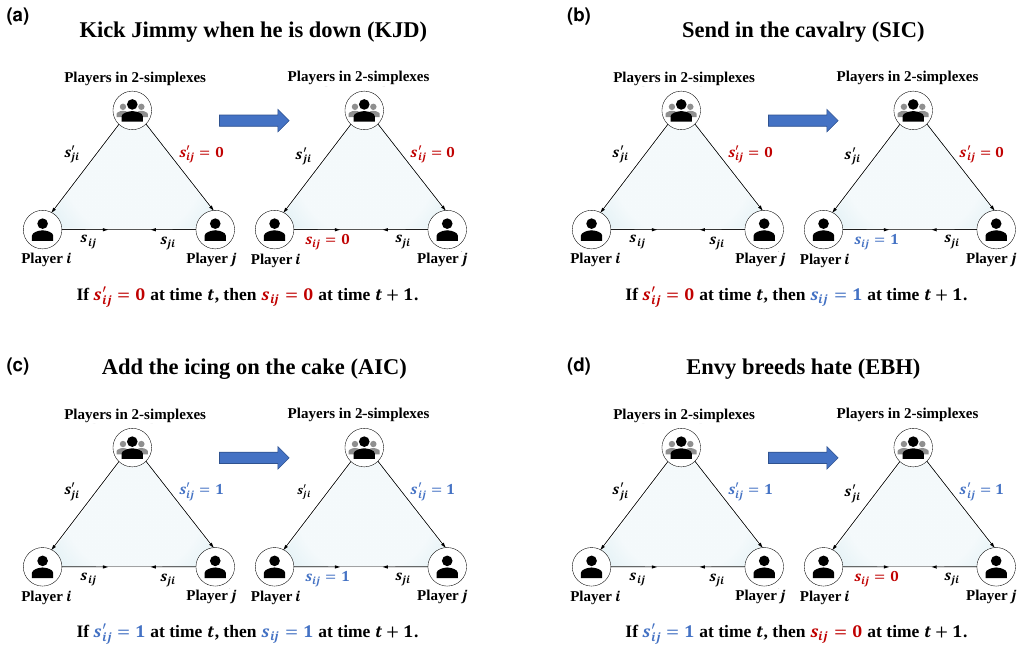}
\caption{\label{Fig:GraphIllu}Illustrations of different strategy updating mechanisms according to social norms: (a) “kick Jimmy when he is down” (KJD), (b) “send in the cavalry” (SIA), (c) “add the icing on the cake” (AIC), and (d) “envy breeds hate” (EBH).}
\end{figure*}

Next, we introduce four social norms and try to figure out if one of them prevails the others and eventually promotes the evolution of cooperation on simplicial complexes. Instead of assuming that each player adopts the same updating strategy and imitates $s^*_{ij}$ with no reference to the collective strategy $s'_{ij}$ at any time, social norms specify how players heterogeneously update $s_{ij}$ when $s'_{ij}$ is considered. The updating strategy of $s_{ij}$ can be decoded into the following $4 \times 3$ matrix:
\begin{equation}\label{Eq:StategyUpdate:Matrix}
    \bordermatrix{%
      & 1 & 0 & \emptyset \cr
    1 & s^*_{ij} & 0 & s^*_{ij}\cr
    2 & s^*_{ij} & 1 &  s^*_{ij}\cr
    3 & 1 & s^*_{ij} & s^*_{ij}\cr
    4 & 0 & s^*_{ij} & s^*_{ij}\cr
    }
\end{equation}
where each row represents player $i$'s social norm $\phi_i$ and each column denotes if collective strategies $s'_{ij} = 1$, $s'_{ij} = 0$, or $s'_{ij} = \emptyset$. The entries in this matrix are then the updating strategies of player $i$ against player $j$ according to the given $\phi_i$ and $s'_{ij}$. As is intuitively illustrated in this matrix, player $i$ with $\phi_i = 1$ is determined to defect player $j$ if the collective strategy is defection ($s'_{ij} = 0$) while cooperates with him under the same circumstance if $\phi_i = 2$. Therefore, we respectively name $\phi_i \in \{1, 2, 3, 4\}$ as “kick Jimmy when he is down” (KJD), “send in the cavalry” (SIC), “add the icing on the cake” (AIC), and “envy breeds hate” (EBH). We also categorize SIC and AIC as prosocial norms while KJD and EBH are antisocial ones. A graphic illustration of these updating mechanisms can be found in Fig.~\ref{Fig:GraphIllu}.

Once social norms are considered in Model II, a third step to update each player's social norm $\phi_i$ will be added before an evolutionary round ends. Each player imitates the social norm of $j^{*}$ with probability $W_{\phi_i \leftarrow \phi_{j^{*}}}$:
\begin{equation}\label{Eq:SocialNormUpdate}
    W_{\phi_i \leftarrow \phi_{j^{*}}} = \frac{1}{1 + \exp[(\Pi_i - \Pi_{j^{*}})/\kappa]}
\end{equation}

\subsection{Model III: Evolutionary games in a noisy environment with private information}\label{Sec:NoisyModel}
Embedded in the models we proposed in Section \ref{Sec:BenchmarkModel} and \ref{Sec:SNModel}, there hides an assumption that every player $i$ has a complete information set on each of his neighbor $j$ in $N_i$ as well as a corresponding collective strategy $s'_{ij}$ to take for reference. However, two strong assumptions are also included with respect to neighbor $j$ and collective strategy $s'_{ij}$: (1) every player $i$ has complete information on all the strategies and payoffs of his neighbor $j$, and (2) $i$ can clearly see the strategies $s_{kj}$ from his neighbors $k \in C^2_{ij}$ towards the target player $j$. These assumptions may sometimes make no sense when noise exists to distort the actual strategies taken by players \cite{Schmid-Ekbatani-Hilbe-Chatterjee-2023-NatCommun}, or when players disagree with the individual assessment due to the private or incomplete information \cite{Hilbe-Schmid-Tkadlec-Chatterjee-Nowak-2018-ProcNatlAcadSciUSA}. Henceforth, we introduce two parameters, $q$ and $\epsilon$, to capture the private information in a noisy environment. Every player still takes the first step to play evolutionary games with all of his neighbors in $N_i$ and calculates payoffs on simplicial complexes for each evolutionary round. After that, when players start to change their strategies towards neighbors, they will be restricted to a limited neighbourhood $\tilde{N}_i \subseteq N_i$. Each player $j \in N_i$ will be chosen with a probability $q$ and constitute a sub-neighbourhood $\tilde{N}_i$. Then, $\tilde{C}^2_{i} \subseteq \tilde{N}_i$ and $\tilde{C}^2_{ij} = \tilde{C}^2_{i} \cap C^2_j$ are defined as neighbors having at least one common 2-simplex relation with $i$ in $\tilde{N}_i$ and common neighbors who are in at least one common 2-simplex with both player $i$ and $j$. Restrained by the private information in $\tilde{N}_i$, player $i$ can only see a limited collective strategy $\tilde{s}'_{ij}$ and adopt a limited potential strategy $\tilde{s}^*_{ij}$ defined as follows:
\begin{equation}\label{Eq:CollectiveStrategyNoisy}
    \tilde{s}'_{ij} =
    \begin{cases}
    1, & \text{if } \frac{1}{\tilde{k}^C_{ij}} \sum_{k \in \tilde{C}^2_{ij}} \tilde{s}^{i}_{kj} \geq \theta \text{ and } \tilde{C}^2_{ij} \neq \emptyset \\
    0, & \text{if } \frac{1}{\tilde{k}^C_{ij}} \sum_{k \in \tilde{C}^2_{ij}} \tilde{s}^{i}_{kj} < \theta \text{ and } \tilde{C}^2_{ij} \neq \emptyset \\
    \emptyset, & \text{if } \tilde{C}^2_{ij} = \emptyset
    \end{cases}
\end{equation}
\begin{equation}\label{Eq:PotentialstrategyNoisy}
    \tilde{s}^*_{ij} =
    \begin{cases}
    s_{\tilde{j}^{*}i}, & \text{if } \tilde{N}_i \neq \emptyset \text{ ,with probability } \frac{1} {{1 + \exp[(\Pi_i - \Pi_{\tilde{j}^{*}})/\kappa]}} \\
    s_{ij}, & \text{otherwise}
    \end{cases}
\end{equation}
where $\tilde{k}^C_{ij}$ is the number of common players in $\tilde{C}^2_{ij}$. $\tilde{j}^{*}$ is the player with the highest payoff in the neighbourhood $\tilde{N}_i$. $\tilde{s}^{i}_{kj}$ $(k \in \tilde{C}^2_{ij})$ is a distorted strategy adopted by player $k$ towards $j$ in the view of player $i$. The distortion is subject to the noise that player $i$ mistakes ${s}_{kj}$ from cooperation to defection or vice versa with probability $\epsilon$. Otherwise, no distortion happens ($\tilde{s}^{i}_{kj} = {s}_{kj}$).

Given the private information in $\tilde{N}_i$ and the noise that distorts ${s}_{kj}$ $(k \in \tilde{C}^2_{ij})$, each player $i$ will then change his strategy $s_{ij}$ towards neighbor $j$ $(j \in N_i)$. If $\tilde{N}_i = \emptyset$, player $i$ will not change $s_{ij}$ since there is no neighbor for reference. Otherwise, $s_{ij}$ will be updated according to a similar $4 \times 3$ matrix with all entries $s^*_{ij}$ substituted by $\tilde{s}^*_{ij}$:

\begin{equation}\label{Eq:StategyUpdateNoisy:Matrix}
    \bordermatrix{%
      & 1 & 0 & \emptyset \cr
    1 & \tilde{s}^*_{ij} & 0 & \tilde{s}^*_{ij}\cr
    2 & \tilde{s}^*_{ij} & 1 &  \tilde{s}^*_{ij}\cr
    3 & 1 & \tilde{s}^*_{ij} & \tilde{s}^*_{ij}\cr
    4 & 0 & \tilde{s}^*_{ij} & \tilde{s}^*_{ij}\cr
    }
\end{equation}
where each row and column corresponds to $\phi_i$ and $\tilde{s}'_{ij}$, respectively.

After synchronously updating every $s_{ij}$ $(i \in N, j \in \tilde{N}_i)$, $\phi_i$ will be updated with a similar Fermi function if $\tilde{N}_i \neq \emptyset$ or, otherwise, remain unchanged:

\begin{equation}\label{Eq:SocialNormUpdateNoisy}
    W_{\phi_i \leftarrow \phi_{\tilde{j}^{*}}} =
    \frac{1}{1 + \exp[(\Pi_i - \Pi_{\tilde{j}^{*}})/\kappa]}
\end{equation}

\begin{figure*}
\centering
\includegraphics[width=0.8\linewidth]{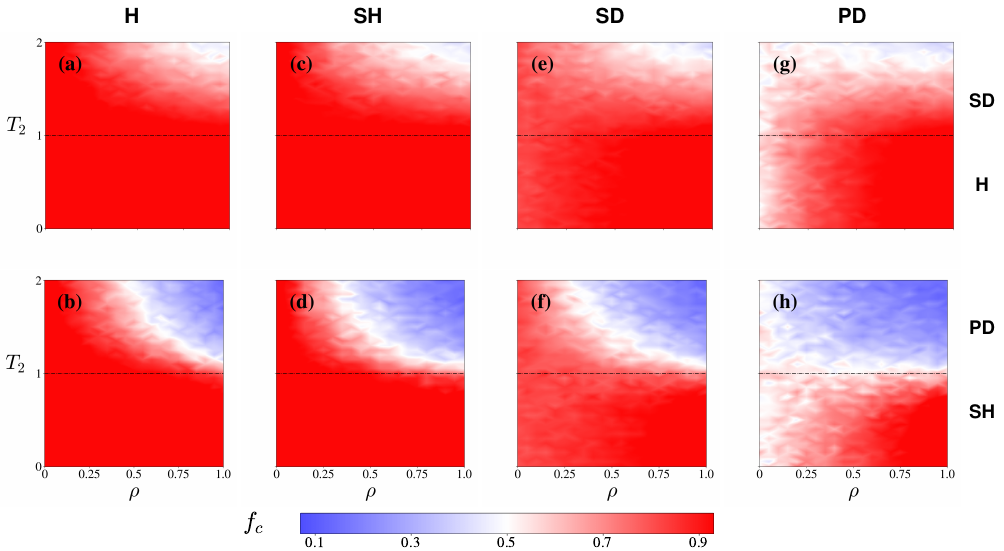}
\caption{\label{Fig:Results:Model1}Contour plots of averaged cooperative strategy fraction $f_c$ in Model I, where all players take the best imitation from their neighbourhood $N_i$ (see Section \ref{Sec:BenchmarkModel} for definition), as a function of 2-simplices fraction $\rho$ and payoff parameter $T_2$. Other parameters are set as follows: the first column defines Harmony game (H, $T_1 = T_3 = 0.8, S_1 = S_3 = 0.2$); the second column defines Stag hunt game (SH, $T_1 = T_3 = 0.8, S_1 = S_3 = -0.2$); the third column defines Snowdrift game (SD, $T_1 = T_3 = 1.2, S_1 = S_3 = 0.2$); the fourth column defines Prisoner's dilemma (PD, $T_1 = T_3 = 1.2, S_1 = S_3 = -0.2$). Top labels indicate different types of Game 1 and Game 3 in their respective columns. On the other hand, each sub-figure has been divided into 2 parts by a horizontal dashed line ($T_2 = 1$), where labels on the right side depict the corresponding type of Game 2 in each sub-panel. We set $S_2 = 0.5$ ($S_2 = -0.5$) for upper (lower) panels.}
\end{figure*}

We set Fermi temperature $\kappa = 0.001$ in Eq.~\ref{Eq:SocialNormUpdate}\&\ref{Eq:SocialNormUpdateNoisy}. Every player synchronously updates all of his strategies $s_{ij}$ and social norm $\phi_i$. In terms of the interactions between players, we consider the fraction of cooperative strategy $f_c$:
\begin{equation}\label{Eq:CoopFraction}
    f_c = \frac{\sum\limits_{i \in N}\sum\limits_{j \in N_i}s_{ij}}{\sum\limits_{i \in N} k_i}
\end{equation}

To dive deep into the evolution of cooperation on simplicial complexes, we also record the fraction of players that adopt social norm $\phi_i$ as $f_{n_i}$, where $i \in \{1,2,3,4\}$ and $j \in \{1,2,\cdots,N\}$:
\begin{equation}\label{Eq:SNFraction}
    f_{n_i} = \frac{\# \{\phi_j | \phi_j = i \}}{N}
\end{equation}

In our simulations, the maximum number of Monte Carlo round is set to be 10000, depending on how quickly $f_c$ stabilizes. With respect to each combination of parameters, $f_c$ and $f_{n_i}$ will be counted in the last 500 steps, where these indicators are in their steady states. To ensure the robustness of our findings and further reduce experimental errors caused by noise, we independently run 5 iterations and average the results. 

\begin{figure*}
\centering
\includegraphics[width=0.8\linewidth]{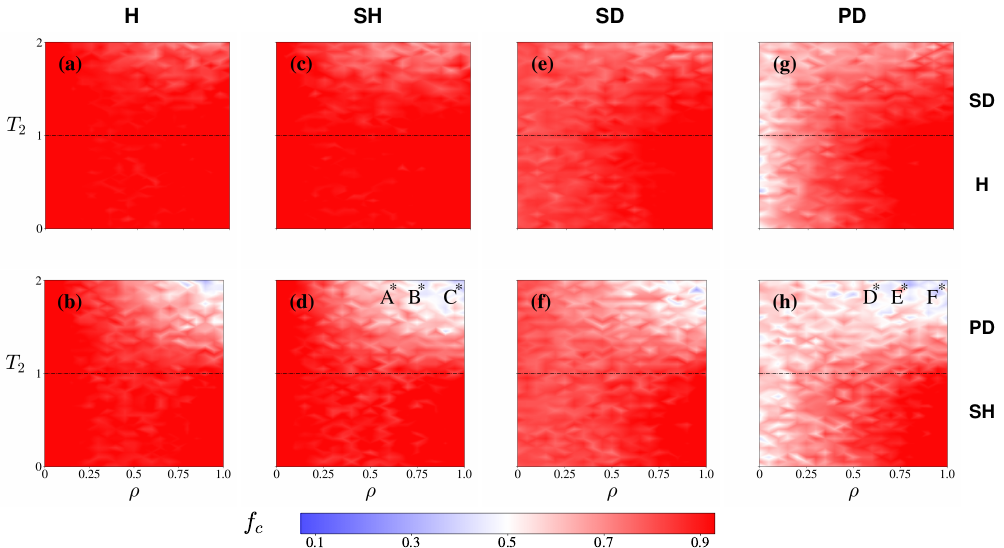}
\caption{\label{Fig:Results:Model2}Contour plots of averaged cooperative strategy fraction $f_c$ in Model II, where social norms are considered with no noise or private information (see Section \ref{Sec:SNModel} for definition), as a function of 2-simplices fraction $\rho$ and payoff parameter $T_2$. Other parameters are set as follows: the first column defines Harmony game (H, $T_1 = T_3 = 0.8, S_1 = S_3 = 0.2$); the second column defines Stag hunt game (SH, $T_1 = T_3 = 0.8, S_1 = S_3 = -0.2$); the third column defines Snowdrift game (SD, $T_1 = T_3 = 1.2, S_1 = S_3 = 0.2$); the fourth column defines Prisoner's dilemma (PD, $T_1 = T_3 = 1.2, S_1 = S_3 = -0.2$). Top labels indicate different types of Game 1 and Game 3 in their respective columns. On the other hand, each sub-figure has been divided into 2 parts by a horizontal dashed line ($T_2 = 1$), where labels on the right side depict the corresponding type of Game 2 in each sub-panel. We set $S_2 = 0.5$ ($S_2 = -0.5$) for upper (lower) panels. Stars in sub-figure (d) and (h) indicate the parameters that are used to plot Fig.~\ref{Fig:Results:CoopDynamics}, with labels representing the specific panel. All stars lie at $T_2 = 1.95$, while $\rho = 0.6$, $0.75$, and $0.95$ for A and D, B and E, as well as C and F, respectively.}
\end{figure*}

\section{Results and discussion}\label{Sec:Results}
\subsection{Social norms can promote cooperation in different competitions of games}
To simplify the parameters that will influence the evolutionary dynamics, we follow Ref.\cite{Guo-Jia-SendinaNadal-Zhang-Wang-Li-AlfaroBittner-Moreno-Boccaletti-2021-ChaosSolitonsFractals} and bind Game 1 and Game 3 to be the same type (i.e., $T_1 = T_3$ and $S_1 = S_3$). With various parameters of Game 2 ($T_2$ and $S_2$) and fractions of 2-simplices ($\rho$), competitions between different Nash equilibria can be well illustrated in a heterogeneous neighbourhood (i.e., a mix of pairwise interactions on 1-simplices and three-body interactions on 2-simplices). Three-body interactions become more prevalent as $\rho$ increases.

Fig.~\ref{Fig:Results:Model1} depicts the average fraction of cooperative strategies $f_c$ in Model I, where all combinations of games in pairwise interactions and three-body interactions are considered. Labels at the top of each column depict the respective game type of Game 1 and Game 3 (from left to right: Harmony game, Stag hunt game, Snowdrift game, and Prisoner's dilemma). $f_c$ is counted in the last 500 rounds and averaged across 5 iterations, where the system is at its asymptotic state. To cover all the competitions between games, we set $S_2 = 0.5$ for the upper panel, where the corresponding type of Game 2 is labeled on the right side of each sub-panel. For illustration, Game 2 is a Harmony game if $0 \leq T_2 < 1$ or a Snowdrift game if $1 < T_2 \leq 2$. The lower panel is analogous, where we set $S_2 = -0.5$.

\begin{figure*}
\centering
\includegraphics[width=0.8\linewidth]{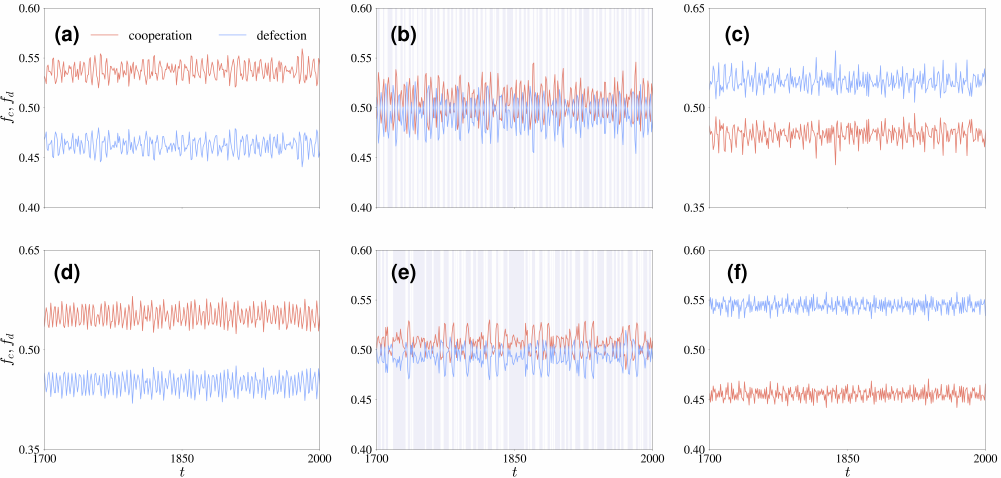}
\caption{\label{Fig:Results:CoopDynamics}Evolutionary dynamics of averaged cooperative (defective) strategy fraction $f_c$ ($f_d$) for different combinations of $T2$, $S_2$ and $\rho$ in Model II. Sub-figures in the upper (lower) row correspond to the circumstances that Game 1 and Game 3 are fixed to be a Stag hunt game (Prisoner's dilemma), with Game 2 kept as a Prisoner's dilemma. Labels in the upper left corner of each sub-figure are mapped to the capital letters and correspond to starred parameter choices in Fig.~\ref{Fig:Results:Model2}. Shaded areas in sub-figure (b) and (e) represent the time steps when cooperation prevails defection.}
\end{figure*}

As is clearly illustrated in Fig.~\ref{Fig:Results:Model1}, cooperation prevails when Game 1 and Game 3 are both Harmony games (i.e., the first column) with respect to low levels of $\rho$, indicating that Nash equilibrium of CC in Game 1 and Game 3 plays a dominant role regardless of the Game 2 type. However, when $\rho$ increases, the fraction of 2-simplices will be enhanced and each player will be engaged in a mix of pairwise interactions and three-body interactions, where different dilemmas compete with each other. According to the first column of Fig.~\ref{Fig:Results:Model1}, the increasing $\rho$ maintains the emergence of cooperation when Game 2 is a Harmony game ($T_2 < 1, S_2 > 0$) or a Stag hunt game ($T_2 < 1, S_2 < 0$), since the Nash equilibria in these games favor cooperation in one way or another. On the other hand, $f_c$ decreases as $\rho$ increases when Game 2 is either a Snowdrift game ($T_2 > 1, S_2 > 0$) or a Prisoner's dilemma ($T_2 > 1, S_2 < 0$). Since the Nash equilibria for SD and PD are CD/DC and DD, respectively, the fraction of cooperative strategy in Fig.~\ref{Fig:Results:Model1} (a) is much larger than that in Fig.~\ref{Fig:Results:Model1} (b) when $\rho \rightarrow 1$ and $T_2 > 1$. Similar results can be also found in the second, third and fourth column of Fig.~\ref{Fig:Results:Model1}, where Game 1 and Game 3 are SH, SD, and PD, respectively. Notably, $f_c$ always reaches a high value when $T_2 < 1$ (i.e., when Game 2 is a Harmony game or a Stag hunt game), especially as $\rho$ increases. This result indicates that the prevalence of 2-simplex interactions may boost the emergence of cooperation if these interactions favour cooperation. However, they can also be double-edged swords, which hinder cooperation when $T_2 > 1$ and $S_2 < 0$. According to Fig.~\ref{Fig:Results:Model1} (b), (d), (f), and (g), $f_c$ always decreases when Game 2 is a Prisoner's dilemma. A clear transition towards defection appears due to the prevalent Nash equilibrium of DD as $\rho$ increases.

\begin{figure*}
\centering
\includegraphics[width=0.8\linewidth]{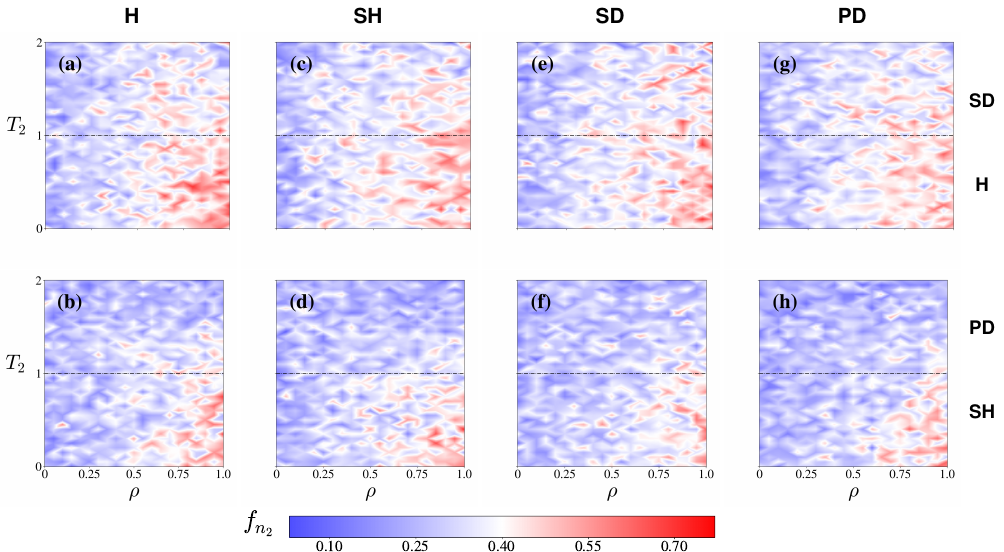}
\caption{\label{Fig:Results:SocialNorm2}The fraction of players who adopt SIC ($f_{n_2}$) at the steady state as a function of parameter $\rho$ and $T_2$ for different game competitions in Model II. Top labels indicate different types of Game 1 and Game 3 in their respective columns: the first column defines Harmony game (H, $T_1 = T_3 = 0.8, S_1 = S_3 = 0.2$); the second column defines Stag hunt game (SH, $T_1 = T_3 = 0.8, S_1 = S_3 = -0.2$); the third column defines Snowdrift game (SD, $T_1 = T_3 = 1.2, S_1 = S_3 = 0.2$); the fourth column defines Prisoner's dilemma (PD, $T_1 = T_3 = 1.2, S_1 = S_3 = -0.2$). Each sub-figure has been divided into 2 parts by a horizontal dashed line ($T_2 = 1$), where labels on the right side depict the corresponding type of Game 2 in each sub-panel. We set $S_2 = 0.5$ ($S_2 = -0.5$) for upper (lower) panels.}
\end{figure*}

Next, we show the results of Model II in Fig.~\ref{Fig:Results:Model2}, where social norms are considered when players update their strategies. Similar trends, that $f_c$ increases with $\rho$ when Game 2 is a Harmony game or a Stag hunt game while decreases when Game 2 is a Prisoner's dilemma, are kept. However, a sharp difference occurs when Fig.~\ref{Fig:Results:Model1} (g) and Fig.~\ref{Fig:Results:Model2} (g) are compared. In the upper part of Fig.~\ref{Fig:Results:Model1} (g), the fraction of cooperative strategies increases when $T_2 \rightarrow 1$  while decreases when $T_2 \rightarrow 2$, which indicates that Nash equilibrium of CD/DC can lead to binary effects on cooperation. Nevertheless, $f_c$ is always increasing as $\rho$ grows in the upper part of Fig.~\ref{Fig:Results:Model2} (g), which implies that the introduction of four social norms, especially SIC and AIC, can provide extra chances for players to cooperate with each other. This argument is further evidenced if we compare all the results in Fig.~\ref{Fig:Results:Model1} and Fig.~\ref{Fig:Results:Model2}, where $f_c$ is always enlarged if social norms are considered in the same combinations of parameter $T_2$ and $\rho$. In order to reach a deeper understanding of how social norms microscopically influence cooperation and prove that our results are not trivial, we further discuss the transitions of strategies and fractions of social norms in the next subsection.

\subsection{Transitions of strategies and fractions of social norms in 2-simplex interactions}
To better explain the evolutionary dynamics when social norms are considered in 2-simplex interactions, we firstly analyze the fraction trait of cooperative strategies $f_c$ and define $f_d = 1 - f_c$ as the fraction of defection. According to six scenarios where Game 2 is constantly a PD with Game 1 and 3 being a SH in the upper row and a PD in the lower one, Fig.~\ref{Fig:Results:CoopDynamics} illustrates three typical regimes that occur when cooperation evolves: (i) cooperation prevails defection (Fig.~\ref{Fig:Results:CoopDynamics} (a) and (d)); (ii) defection defeats cooperation (Fig.~\ref{Fig:Results:CoopDynamics} (c) and (f)) and (iii) time-varying dominance of cooperation or defection (Fig.~\ref{Fig:Results:CoopDynamics} (b) and (e)). These results, especially those in Fig.~\ref{Fig:Results:CoopDynamics} (b) and (e) where an asymptotic state of alternatively prevailed cooperation and defection exists, strongly evidence that our model can provide players with free options to change their strategies and prove that our findings are not trivial. When social norms are added in our model, time-dependent regime shifts of cooperation and defection can be attached to empirical implications of ever-changing group decisions according to majority rules \cite{Hastie-Kameda-2005-PsycholRev}, as well as close relations between election outcomes and updating social norms \cite{Apffelstaedt-Freundt-Oslislo-2022-JEconBehavOrgan}.

\begin{figure*}
\centering
\includegraphics[width=0.8\linewidth]{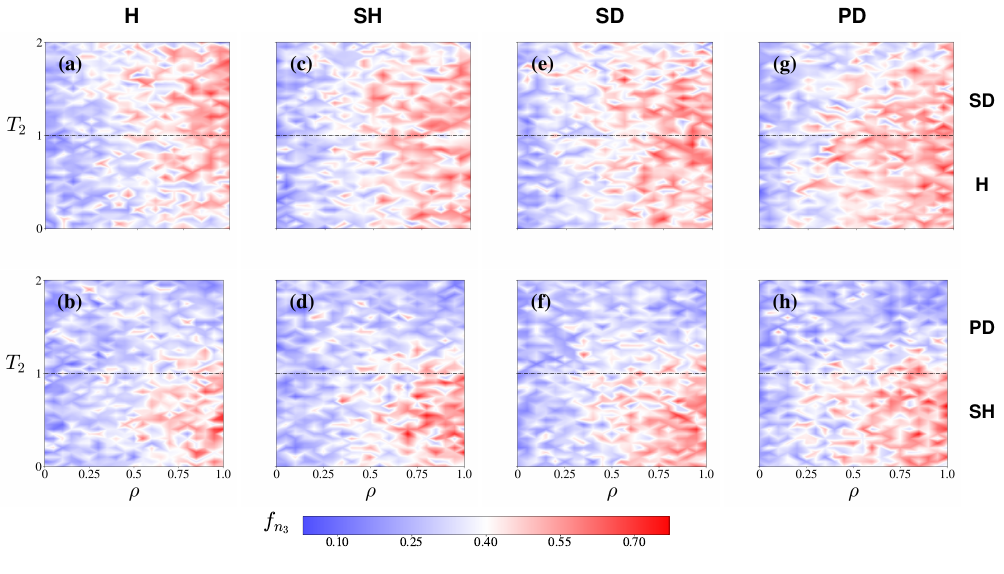}
\caption{\label{Fig:Results:SocialNorm3}The fraction of players who adopt AIC ($f_{n_3}$) at the steady state as a function of parameter $\rho$ and $T_2$ for different game competitions in Model II. Top labels indicate different types of Game 1 and Game 3 in their respective columns: the first column defines Harmony game (H, $T_1 = T_3 = 0.8, S_1 = S_3 = 0.2$); the second column defines Stag hunt game (SH, $T_1 = T_3 = 0.8, S_1 = S_3 = -0.2$); the third column defines Snowdrift game (SD, $T_1 = T_3 = 1.2, S_1 = S_3 = 0.2$); the fourth column defines Prisoner's dilemma (PD, $T_1 = T_3 = 1.2, S_1 = S_3 = -0.2$). Each sub-figure has been divided into 2 parts by a horizontal dashed line ($T_2 = 1$), where labels on the right side depict the corresponding type of Game 2 in each sub-panel. We set $S_2 = 0.5$ ($S_2 = -0.5$) for upper (lower) panels.}
\end{figure*}

Next, we analyze how social norms can change and make heterogeneous contributions to the emergence of cooperation from a microscopic perspective. Since SIC and AIC are two representative norms that have potential to promote cooperation when collective strategies in the neighbourhood are given, we put emphasis on these norms and depict the faction of players who adopt them as a function of parameter $\rho$ and $T_2$ in Fig.~\ref{Fig:Results:SocialNorm2} and Fig.~\ref{Fig:Results:SocialNorm3}, respectively. When $\rho$ increases and three-body interactions on 2-simplices prevail in the whole network, it can be clearly observed that SIC and AIC are adopted more frequently if Game 2 is a Harmony game, a Stag hunt game, or a Snowdrift game. As is intuitively evidenced by darker areas in Fig.~\ref{Fig:Results:SocialNorm3}, AIC reaches a higher popularity when Game 2 is set to be a game that is more or less in favour of cooperation (i.e., H, SH, or SD). Taking SH as an example, AIC can trigger the diffusion of cooperation as long as the collective strategy towards a player is to cooperate. Since one of the Nash equilibria in SH is CC, cooperation then emerges and stimulates an increasing fraction of players to adopt AIC. In another aspect, the emergence of SIC and AIC can explain the increasing $f_c$ when Game 2 is a Snowdrift game, as displayed in Fig.~\ref{Fig:Results:Model2}. Due to the Nash equilibria of CD/DC in SD, SIC provides an opportunity for players to cooperate ($s_{ij} = 1$) with those who occupy positive direct strategies ($s_{ji} = 1$) but receive negative feedbacks ($s'_{ij} = 0$). This mutual cooperation can, therefore, be further reinforced when AIC takes its part in the evolutionary dynamics, which eventually contributes to the emergence of cooperation on 2-simplices from a holistic perspective. However, neither SIC nor AIC can be drastically promoted when $T_2 \rightarrow 2$ and $S_2 = -0.5$, where Game 2 is a PD and gets dominant as $\rho$ grows. Only when $T_2 \rightarrow 1$ and $\rho \rightarrow 1$ can we find subtle increment of $f_{n_2}$ and $f_{n_3}$, which facilitate some sort of resistance to defection and contribute to the emergence of cooperation while comparing Fig.\ref{Fig:Results:Model2} (h) with Fig.~\ref{Fig:Results:Model1} (h).

\begin{figure*}
\centering
\includegraphics[width=0.8\linewidth]{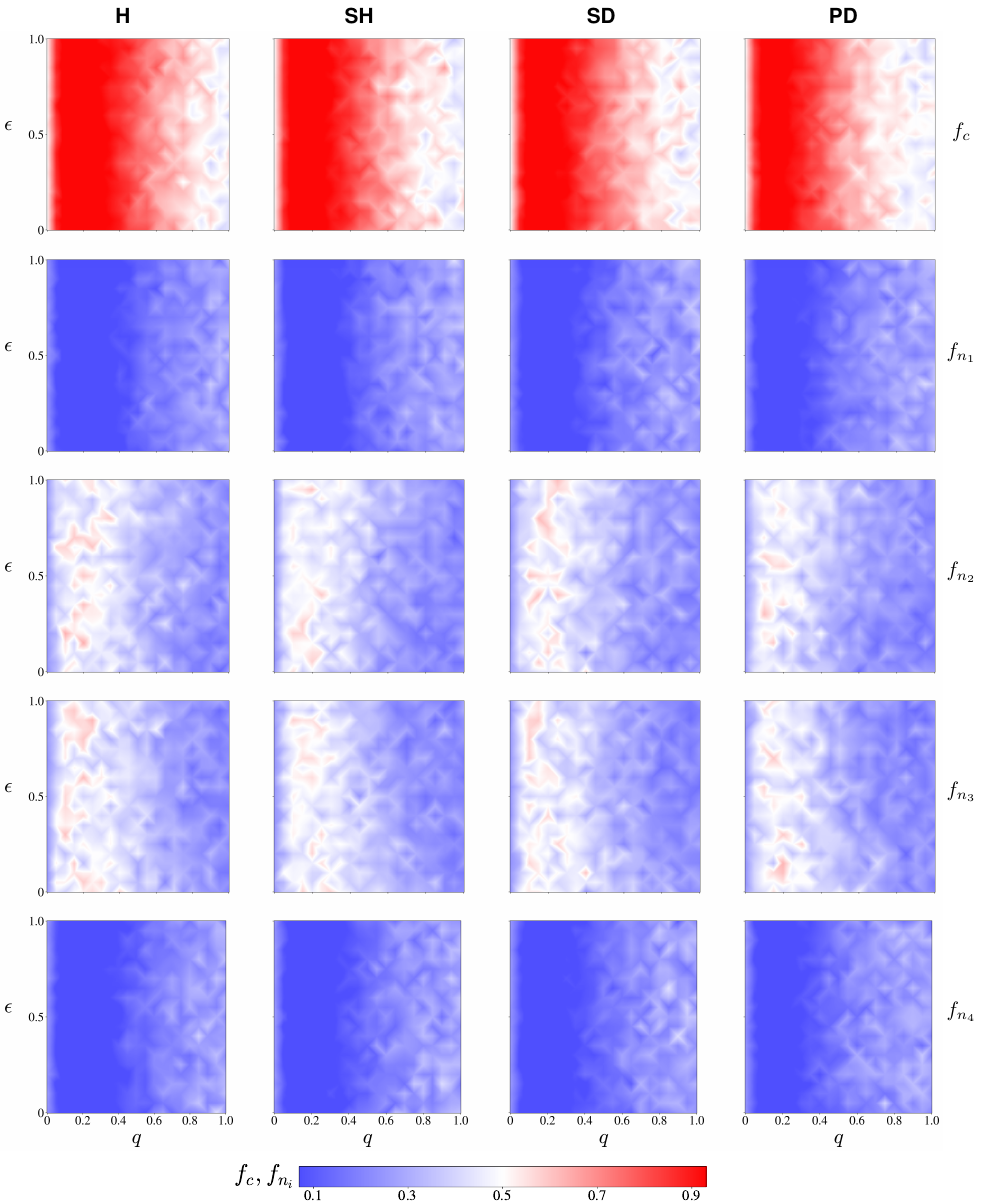}
\caption{\label{Fig:Results:NoisyEnvPD}The fraction of averaged cooperative strategy $f_c$ and fractions of players who adopt KJD ($f_{n_1}$), SIC ($f_{n_2}$), AIC ($f_{n_3}$), and EBH ($f_{n_4}$) as functions of parameter $q$ and $\epsilon$ when Game 2 is set to be a PD ($T_2 = 2$, $S_2 = -0.5$) in Model III (see Section \ref{Sec:NoisyModel} for definition). Top labels indicate different types of Game 1 and Game 3 in their respective columns: the first column defines Harmony game (H, $T_1 = T_3 = 0.8, S_1 = S_3 = 0.2$); the second column defines Stag hunt game (SH, $T_1 = T_3 = 0.8, S_1 = S_3 = -0.2$); the third column defines Snowdrift game (SD, $T_1 = T_3 = 1.2, S_1 = S_3 = 0.2$); the fourth column defines Prisoner's dilemma (PD, $T_1 = T_3 = 1.2, S_1 = S_3 = -0.2$). Right labels depict respective indicators for each row. Other parameters are set as follows: $\rho = 1$, $\theta = 0.5$, and $\kappa = 0.001$.}
\end{figure*}

\subsection{Cooperation and social norms on 2-simplices in a noisy and private environment}
As is introduced in Section \ref{Sec:Introduction} that a noisy and private environment can have potential effects on players' evaluation towards each other, which further influences collective strategies $s'_{ij}$ and activates updates of $s_{ij}$ according to different social norms, we, therefore, analyze how cooperation and social norms will evolve when noise and information privacy are considered. To capture the intensity of noise and information privacy, we introduce two parameters, $\epsilon$ and $q$, in Model III. A larger $\epsilon$ indicates a noisier environment while information in neighbourhoods are much more private when $q$ decreases.

Fig.~\ref{Fig:Results:NoisyEnvPD} illustrates contour plots of $f_c$ and $f_{n_1} - f_{n_4}$ with the $\epsilon-q$ phase diagrams, where Game 2 is fixed to be a PD. The parameter settings and corresponding labels for each sub-figure have been detailedly shown in the caption of Fig.~\ref{Fig:Results:NoisyEnvPD}. It can be clearly observed that: (i) $\epsilon$ seems to have little influence on $f_c$ and $f_{n_1} - f_{n_4}$ since no extraordinary change happens in a vertical way when $q$ is given. This result makes sense because a majority rule is applied in our model and the updates of strategies should rely on collective strategies shaped by neighbors, instead of a specific judgement influenced by noise; (ii) $f_{n_1}$ and $f_{n_4}$ share a common trend that they reach a higher level when $q \rightarrow 1$; (iii) $f_{n_2}$ and $f_{n_3}$ can be kept at a relatively higher level when information is private to some extend ($q \approx 0.2$). Despite a strong advocacy for mutual defection in PD, information privacy filters defection from others and provides potential chances to facilitate social norms that breeds cooperation (i.e., SIC and AIC). Taking all these findings together, it is natural to explain that (iv) $f_c$ increases when $q$ is at a moderate level but decreases if $q$ continues to grow, which is comprehensively influenced by the combination of four social norms.

\section{Discussion}\label{Sec:Conclusions}
Prosocial behavior is common in everyday life while being at odd with the traditional view of selfishness in natural selection \cite{Hamilton-1970-Nature}. When mechanisms like kin selection and direct reciprocity are considered, a substantial degree of cooperation becomes evident. To further understand the widespread nature of cooperation, significant focus has been placed on indirect reciprocity. In this context, the way individuals assess others and modify their strategies accordingly is critical in the evolution of cooperation.

In our research, we have introduced three innovative models to explore the pivotal role of cooperation in the evolution of collective strategies influenced by neighbors and social norms in higher-order networks. We investigated how evaluations made by third parties in settings with both pairwise and three-body interactions impacting strategy updates through four social norms (KJD, SIC, AIC, and EBH), aiding the rise of cooperation. Our study examines how these norms promote cooperation in various social dilemmas with different Nash equilibria and sustain cooperation in environments characterized by noise and privacy, particularly when mutual defection is otherwise favored. Our findings indicate that incorporating social norms consistently boosts cooperation, regardless of the type of dilemma or the nature of interactions, whether pairwise or higher-order.

We also provide insights into why social norms evolve and support cooperation in higher-order networks. Prosocial norms like SIC and AIC are particularly effective in scenarios that favor cooperation. Even in tit-for-tat equilibria (CD/DC) within three-body interactions, the combined influence of collective strategies and social norms can foster mutual cooperation. Our models demonstrate that a slight increase in SIC or AIC adoption can lead to cooperation even when mutual defection is the Nash equilibrium, provided the temptation to defect is moderate. These models replicate three typical regimes, reflecting real-world scenarios where the influence of social norms can be critical in decisions like elections and group choices, especially when majority rules are applied.

We further explore how social norms encourage cooperation on 2-simplices in noisy and private settings. While noise has minimal impact due to our majority rule for collective strategies, the level of information privacy becomes crucial, especially in Prisoner's Dilemma situations on 2-simplices. A moderate level of privacy helps maintain prosocial norms and limit antisocial ones, leading to increased cooperation. This suggests that an optimal level of privacy can deter defectors and support cooperative strategies. These findings highlight the significance of privacy protection in social networks to promote prosocial norms and harmony.

Future extensions of our work could include introducing preferential attachment in mixed interaction structures, varying the constraints between different games, and examining the initial distributions of social norms. These directions could provide a yet richer understanding of the dynamics at games in the evolution of cooperation and social norms.

\begin{acknowledgments}
Y.M. acknowledges partial support from the China Scholar Council State Scholarship Fund (No. 202306740033). Z.J. acknowledges partial support from the National Natural Science Foundation of China (No. 72131007 and No. 72171083), the National Social Science Foundation (Youth Program) of China (No. 20CJY020), and the Fundamental Research Funds for the Central Universities. F.F. acknowledges partial support from the China Scholar Council State Scholarship Fund (No. 202306830151). M.P. was supported by the Slovenian Research and Innovation Agency (Javna agencija za znanstvenoraziskovalno in inovacijsko dejavnost Republike Slovenije) (Grant Nos. P1-0403 and N1-0232). S.B. acknowledges support from the project n.PGR01177  "Systems with higher order interactions and fractional derivatives for applications to AI and high performance computing" of the Italian Ministry of Foreign Affairs and International Cooperation.
\end{acknowledgments}

\bibliography{SocialNormsSC_MP_SB.bib}

\end{document}